\documentclass[preprint,11pt,3p]{elsarticle}

\usepackage{amssymb}
\usepackage{graphicx}
\usepackage{amsmath}
\usepackage{amssymb}
\usepackage{tikz}
\usepackage{secdot}
\biboptions{numbers,sort&compress}
\usepackage{tabularx,ragged2e,booktabs,caption}
\usepackage{color}
\usepackage{cite}
\usepackage{float}

\journal{Journal of Alloys and Compounds}
\begin{document}

\begin{frontmatter}
\title{ Identification of phase components in Zr-Ni and Hf-Ni intermetallic compounds; Investigations by perturbed angular correlation spectroscopy and 
first principles calculations
}
\author{S.K. Dey$^{1,2}$}
\ead{skumar.dey@saha.ac.in}
\author{C.C. Dey$^{1,2}$\corref{cor1}}
\cortext[cor1]{corresponding author}
\ead{chandicharan.dey@saha.ac.in}
\author{S. Saha$^{1,2}$}
\ead{satyajit.saha@saha.ac.in}
\author{J. Belo$\check{\text{s}}$evi\'{c}-$\check{\text{C}}$avor$^3$}
\ead{cjeca@vin.bg.ac.rs}
\author{D. Toprek$^3$}
\ead{toprek@vin.bg.ac.rs}
\address{$^1$Saha Institute of Nuclear Physics, 1/AF Bidhannagar, Kolkata-700 064, India}
\address{$^2$ Homi Bhabha National Institute, Anushaktinagar, Mumbai-400 094, India}
\address{$^3$Institute of Nuclear Sciences Vinca, University of Belgrade, P. O. Box 522,\\ 11001 Belgrade, Serbia}

\begin{abstract}

Time-differential perturbed angular correlation (TDPAC) measurements have been
carried out in stoichiometric ZrNi$_3$ and HfNi$_3$ intermetallic compounds using $^{181}$Ta
probe in the temperature range 77-1073 K considering the immense technological applications of Zr-Ni and Hf-Ni intermetallic compounds. In ZrNi$_3$,
four components due to the production of Zr$_2$Ni$_7$, Zr$_8$Ni$_{21}$, Zr$_7$Ni$_{10}$ and ZrNi$_3$ have been
found at room temperature. The HfNi$_3$ sample produces five electric
quadrupole interaction frequencies at room temperature. The phase HfNi$_3$ is strongly produced in stoichiometric sample of HfNi$_3$ where two
non-equivalent Hf sites are found to be present. Besides this phase, 
two other phases due to Hf$_2$Ni$_7$ and Hf$_8$Ni$_{21}$ have been found but, we do not observe any phase due to Hf$_7$Ni$_{10}$. X-ray
diffraction, TEM/energy dispersive X-ray spectroscopy (EDX) and TEM-selected area electron diffraction (SAED) measurements
were used to further characterize the investigated materials
and it was found that these results agree with the TDPAC results.
In order to confirm findings from TDPAC measurements, density functional theory (DFT)
based calculations of electric field gradients (EFG) and asymmetry parameters
at the sites of $^{181}$Ta probe nucleus were performed. Our calculated results are found to be in excellent agreement with the experimental results.

\end{abstract}
\begin{keyword}
A. hydrogen absorbing materials; A. intermetallics; B. mechanical alloying; C. hyperfine interactions; D. perturbed angular correlations, PAC; D. X-ray diffraction;

\end{keyword}

\end{frontmatter}
\section{Introduction}
 
The elements zirconium and hafnium are alloyed with cobalt, nickel,
titanium, palladium etc. to form many intermetallic compounds which have
technological applications. Zirconium-nickel alloys are found to have useful
hydrogen storage properties. It was shown that the compounds
Zr$_8$Ni$_{21}$, Zr$_9$Ni$_{11}$, Zr$_7$Ni$_{10}$, Zr$_2$Ni$_7$, ZrNi are
good hydrogen absorbing materials to form interstitial metal hydrides (MH) which have important application in nickel metal hydride (NiMH) batteries as negative 
electrode material. The electrochemical properties of several Zr-Ni intermetallic compounds
were studied earlier \citep{JMJoubert,Ruiz,FCRuiz,S.O.Salley,Regmi,Young} by different workers. In ZrNi$_3$, catalytic hydrogen activity was
reported by Wright et al. \citep{RBWright}. The hafnium alloyed with nickel, niobium,
and tantalum are also useful and can withstand high
temperature and pressure. Hafnium alloys are useful in
medical implants and devices due to their bio-compatibility
and corrosion resistance \citep{Davidson}. The alloys of Ni-Ti-Hf
exhibit shape memory behavior \citep{Meng}. Intermetallic compounds
of Hf and transition metals (Fe, Co, Pd, Pt) have also
hydrogen storage properties \citep{Baudry}, with high H/M
ratio at room temperature.

Time differential perturbed angular correlation (TDPAC)
or simply PAC is an important nuclear technique to study
the structural properties of compounds that contain hafnium
and zirconium. Effects of a $\gamma$-$\gamma$ angular correlation in a crystalline
environment are measured by this technique through hyperfine
interaction (electric quadrupole and/or magnetic dipole).
In electric quadrupole interaction, the quadrupole
moment of the probe nucleus interacts with the electric
field gradient (EFG) that arises in a crystalline material
with noncubic symmetry due to charge distribution
of the probe environment. In magnetic interaction, the magnetic dipole
moment of the probe nucleus interacts with the internal/external
magnetic field. Using this technique, several studies
in Hf/Zr-Ni systems were carried out earlier to investigate
their EFGs and magnetic properties \citep{Silva,Marszalek,Wodnecka,Poynor,Koteski,Ivanovski,CCDey_physica}. Recently we have
studied the structural properties of (Zr/Hf)$_8$Ni$_{21}$, (Zr/Hf)$_7$Ni$_{10}$
using the PAC technique \citep{skdeyZr8Ni21,skdeyZr7Ni10}. However, we do not find any previous PAC studies in (Zr/Hf)Ni$_3$. In the present
report, attempts
have been made to produce the intermetallic compounds (Zr/Hf)Ni$_3$
by arc melting of the constituent elements taken in stoichiometric
ratios and characterize them by PAC spectroscopy. According to Becle et al. \citep{Becle}, the ZrNi$_3$ was
formed from Zr$_2$Ni$_5$ and Zr$_2$Ni$_7$ at 940$^\circ$C
or below by a peritectoid reaction. The stable phase of ZrNi$_3$ was obtained at room temperature by annealing the sample
at $\sim$860$^\circ$C. They found 
that the phase 
ZrNi$_3$ was not stable and decayed at 940$^\circ$C following 4ZrNi$_3\rightarrow$Zr$_2$Ni$_7$+Zr$_2$Ni$_5$.
The phase ZrNi$_3$, however, was not found by other workers \citep{Smith, Kramer, Kirkpatrick}. J. H. N. Van Vucht \citep{Vucht} also
failed to produce ZrNi$_3$ by replacing
Ti with Zr in TiNi$_3$. From theoretical calculations by density functional theory (DFT), the
compounds ZrNi$_3$ and HfNi$_3$ were found to be highly stable alloys. Values of enthalpies of formation
of these compounds were reported to be -0.36 eV/atom (ZrNi$_3$) and -0.44 eV/atom (HfNi$_3$) \citep{Johannesson}. In order to help identifying the different phases produced
in the investigated samples, the electric field gradients
were calculated by density functional theory (DFT) and
compared with the measured EFGs. The temperature dependent
PAC measurements enabled us to find any structural changes
in the material and give information on the structural stability of the compound.

The ZrNi$_3$ is known to be a hexagonal close-packed compound of the SnNi$_3$ type with space group $P6_3/mmc$. The lattice parameters were reported
 to be $a$=5.309 \r{A} and $c$=4.303 \r{A} \citep{Becle}. Crystal structure of HfNi$_3$ was also reported by L. Bsenko \citep{HfNi3Structure}. It was
 found \citep{HfNi3Structure} that HfNi$_3$ exists in two modifications. A high temperature $\alpha$-HfNi$_3$ phase and a low temperature $\beta$-HfNi$_3$ 
 phase. The crystal parameters for the two phases were reported to be $a$=5.27 \r{A}, $c$=19.2324 \r{A} and $a$=5.2822 \r{A}, $c$=21.3916 \r{A} for
 the $\alpha$ and $\beta$ phases, respectively.

\begin{figure*}[t!]
\begin{center}
\includegraphics[width=0.8\textwidth]{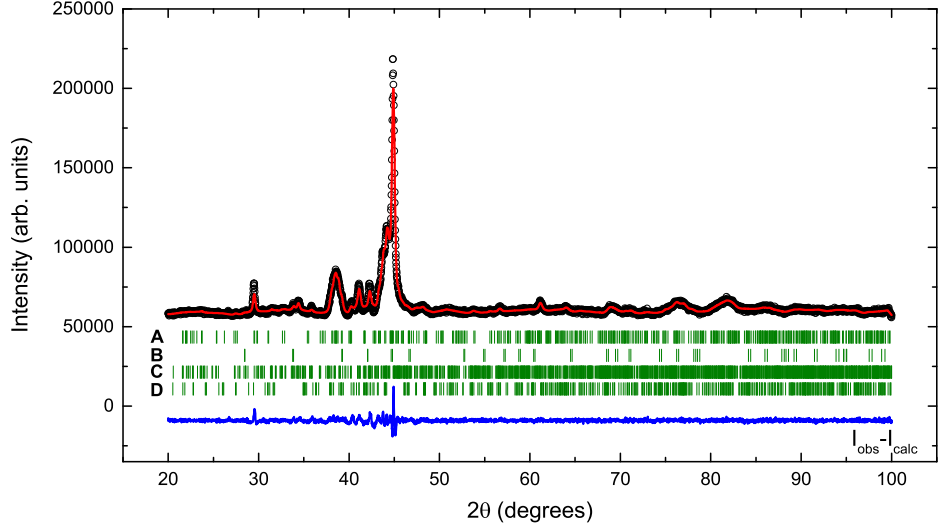}
\end{center}
\captionof{figure}{\small{The background subtracted XRD powder pattern in the stoichiometric sample of ZrNi$_3$. The line represents the fit to the measured data. 
The vertical bars A, B, C and D denote the Bragg angles corresponding to Zr$_2$Ni$_7$, ZrNi$_3$, Zr$_8$Ni$_{21}$ and Zr$_7$Ni$_{10}$, respectively. 
The bottom line shows the difference between the observed and the fitted pattern.}}
\label{ZrNi3_XRD}
\end{figure*}

\begin{figure*}[t!]
\begin{center}
\includegraphics[width=0.9\textwidth]{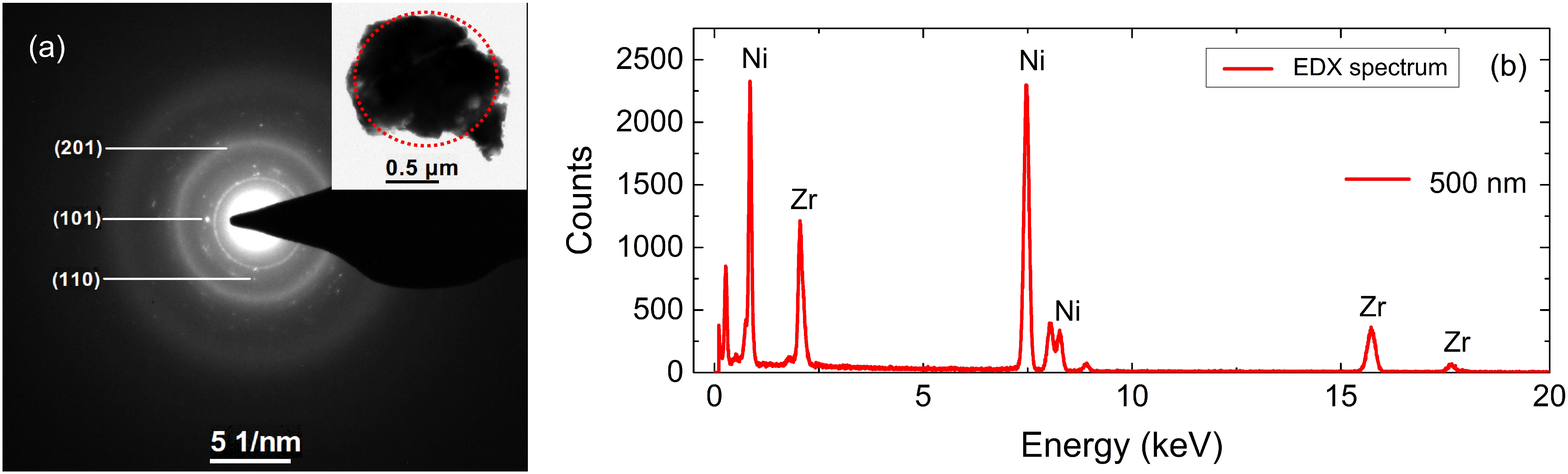}
\end{center}
\captionof{figure}{\small{(a) Selected area electron diffraction pattern from ZrNi$_3$ particle shown in the inset. (b) Energy dispersive X-ray spectrum
from the same particle.}}
\label{ZrNi3_TEM}
\end{figure*}

\begin{figure*}[t!]
\begin{center}
\includegraphics[width=0.7\textwidth]{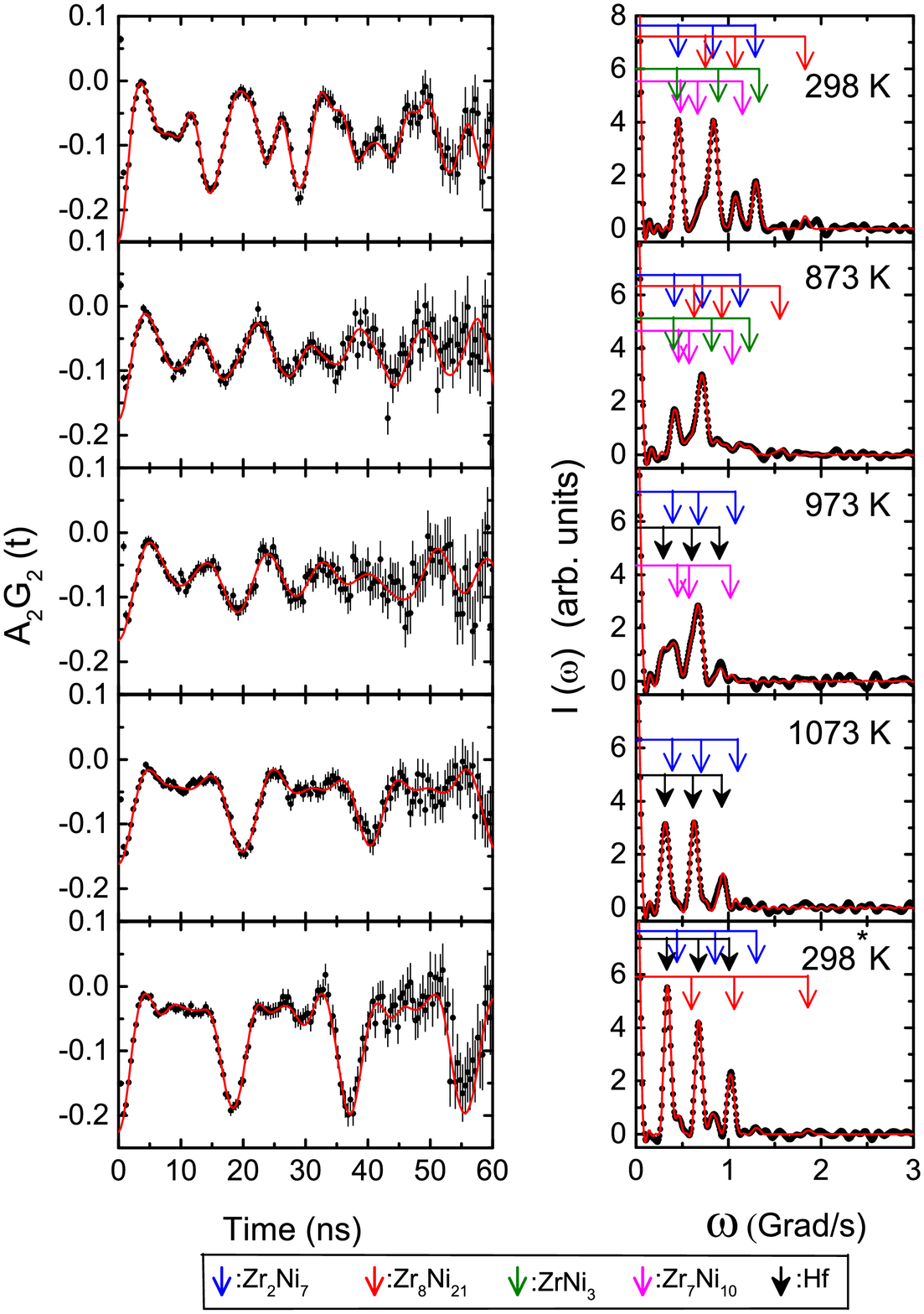}
\end{center}
\captionof{figure}{\small{TDPAC spectra in the stoichiometric sample of ZrNi$_3$ at different
temperature. Left panel
shows the time spectra and the right panel shows
the corresponding Fourier cosine transforms. The PAC spectrum at room temperature designated by 298$^*$ K is taken after the
measurement at 1073 K. A set of three
arrows indicates the three transition frequencies of a particular component.}}
\label{ZrNi3_spectra}
\end{figure*} 

\begin{figure}[h!]
\begin{center}
\includegraphics[width=0.5\textwidth]{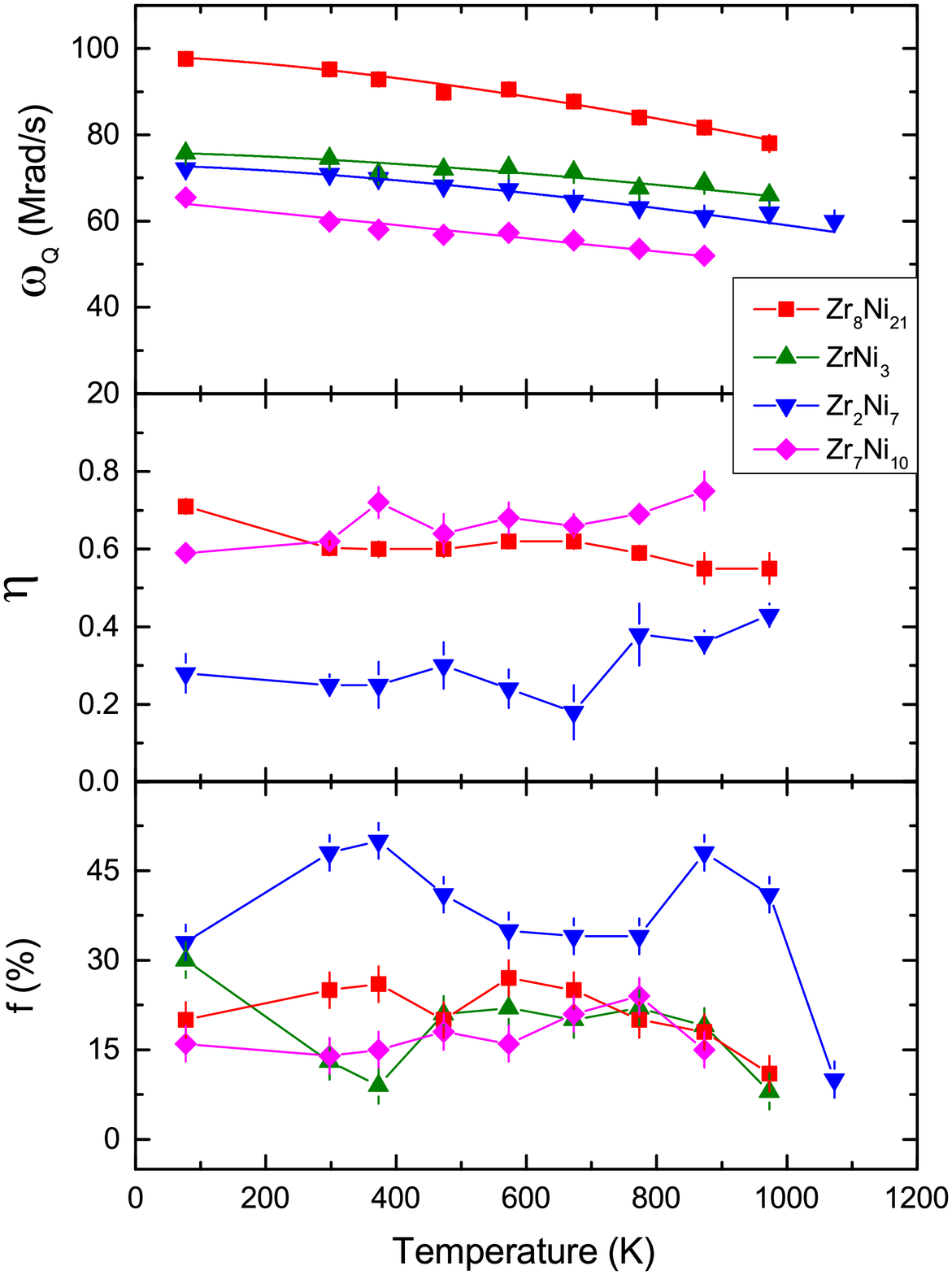}
\end{center}
\captionof{figure}{\small{ Variations of quadrupole frequency ($\omega_Q$), asymmetry parameter ($\eta$) and site fraction $f$(\%)
with temperature for the components of Zr$_2$Ni$_7$, ZrNi$_3$, Zr$_8$Ni$_{21}$ and Zr$_7$Ni$_{10}$.}}
\label{ZrNi3_parameter}
\end{figure}

\section{Experimental details}
To produce the intermetallic compounds ZrNi$_3$ and HfNi$_3$, stoichiometric
amounts of constituent elements procured from M/S Alfa Aesar were taken. The purity 
of the metals used were : Zr-99.2\% (excluding Hf), Hf-99.95\%(excluding Zr) and Ni-99.98\%.
To introduce the $^{181}$Hf probe, each sample was remelted 
by adding an active piece of Hf wire ($\sim$1 mg). Shiny globule samples
were formed
after melting in the arc furnace. Natural Hf ($\sim$ 30\% $^{180}$Hf)
was pre-activated to $^{181}$Hf in Dhruba reactor, Mumbai by thermal neutron capture with a flux $\sim$ 10$^{13}$/cm$^2$/s for 7 days.
Samples 
were then sealed in evacuated quartz tubes to carry out measurements at high temperatures. 
Separate inactive stoichiometric samples of ZrNi$_3$ and HfNi$_3$ were also prepared in similar manners for X-ray diffraction and TEM/energy dispersive X-ray spectroscopy (EDX) measurements.
XRD measurements
were carried out using the Rigaku X-ray diffractometer TTRAX-III and Cu $K_\alpha$ radiation. 
Transmission electron microscopy (TEM) measurements were carried out using
FEI, Tecnai G2 F30, S-Twin microscope equipped with a high angle annular dark-field (HAADF) detector, a
scanning unit and a energy dispersive X-ray spectroscopy (EDX) unit to
perform the scanning transmission electron
microscopy (STEM-HAADF-EDX).

The perturbed angular correlation is a nuclear technique to measure the hyperfine interactions between the nuclear moments
of the probe nucleus and the hyperfine fields present in the investigated sample. The probe $^{181}$Hf emits two successive $\gamma$-rays, 133 and 
482 keV, passing through the 482 keV intermediate level ($T_{1/2}$=10.8 ns) with a spin angular momentum $I$=5/2$^+\hbar$ \citep{Stone}. 
The extra-nuclear 
electric field gradients present in the sample interact with the nuclear quadrupole moment of the intermediate level ($Q$=2.35 b \citep{Stone}).
Due to this interaction, the angular 
correlation of the 133-482 keV $\gamma$-$\gamma$ cascade is perturbed. The perturbation function is given by \citep{Schatz},
    \begin{equation}
  G_2(t)=\Big[S_{20}(\eta) + \sum^{3}_{i=1}S_{2i}(\eta)cos(\omega_it)exp(-\delta\omega_it)exp\big[\frac{-(\omega_i\tau_R)^2}{2}\big]\Big].
 \label{eqn:G22}
 \end{equation}
 The above expression of perturbation function is valid for a polycrystalline sample and for $I$=5/2$^+$ of intermediate state of the probe nucleus. 
 The frequencies $\omega_i$ are the transition frequencies between different $m$-sublevels arising due to hyperfine splitting. A damping of perturbation 
 function (Lorentzian) was considered through the first exponential which can arise due to structural defects in the sample. Here, $\delta$ is the 
 frequency distribution width. The finite time resolution ($\tau_R$) of the coincidence set up was considered through the second exponential. If 
 more than one quadrupole interaction is present in the sample due to the presence of different component
 phases or due to two or more non-equivalent sites of 
 a particular phase, the perturbation function can be written as       
      \begin{equation}
 G_2(t)=\sum_if_iG^{i}_{2}(t)
  \label{eqn:Bhava}
 \end{equation}
 where, $f_i$ is the fraction of the $i$-th component and $G^{i}_2(t)$ is the corresponding perturbation function. 
     A fitting to expression (\ref{eqn:G22}) determines the quadrupole frequency $\omega_Q$ through the measured values of $\omega_1$, $\omega_2$
     and $\omega_3$. The quadrupole frequency is directly related to the electric field
    gradient ($V_{zz}$) through the relation
      \begin{equation}
\omega_Q=\frac{eQV_{zz}}{4I(2I-1)\hbar}.
 \label{eqn:rita}
\end{equation}
    For an axially
    symmetric EFG ($\eta$=0), $\omega_Q$ is related to $\omega_1$, $\omega_2$ and $\omega_3$ by
    $\omega_Q$=$\omega_1$/6=$\omega_2$/12=$\omega_3$/18. 
        The asymmetry parameter is defined as the ratio 
     \begin{equation}
  \eta=\frac{(V_{xx}-V_{yy})}{V_{zz}}
  \label{eqn:newton}
 \end{equation}
    and
    its value lies between 0 and 1. For $\eta\ne$0, this simple relation between $\omega_Q$ and $\omega_i$'s does not hold but, produces a more complex
    relation \citep{Zacate}.
    
 A four detector 
   LaBr$_3$(Ce)-BaF$_2$ set up was used for present TDPAC measurements. The crystal sizes were 38$\times$25 mm$^2$ and 51$\times$51 mm$^2$ for LaBr$_3$(Ce)
   and BaF$_2$, respectively. The 133 keV $\gamma$-rays were detected in the LaBr$_3$(Ce) detector and the
   482 keV $\gamma$-rays were detected in the BaF$_2$ detector. 
   Standard slow-fast
   coincidence assemblies were employed to acquire four coincidence spectra at 180$^o$ and 90$^o$ \citep{pramana}. 
   A typical prompt time resolution (FWHM) of $\sim$800 ps has been obtained for the energy window settings of $^{181}$Ta $\gamma$-rays.
   The perturbation function $G_2$($t$) was obtained from the ratio of coincidence counts at 180$^o$ and 90$^o$. 
     Details on the experimental
     set up and data acquisition  can be found in our earlier report \citep{pramana}.

\begin{table}[h]
\begin{center} 
\captionof{table}{\small{Results of temperature dependent variations of $\omega_Q$ for different components in the stoichiometric samples of ZrNi$_3$ and HfNi$_3$.}}
\scalebox{0.65}{
\begin{tabular}{ l l l l l l l } 
 \hline  \\ [-0.9ex]
Component phases  & $\omega_Q$(0)   &$V_{zz}$(0)  & $\alpha$ & $\beta$ \\ &(Mrad/s) &($\times$10$^{21}$) V/m$^2$ &($\times$10$^{-4}$) K$^{-1}$  &($\times$10$^{-6}$) K$^{-3/2}$     \\ [1.5ex]
 \hline  \\ 
 
Zr$_2$Ni$_7$                     &72.9(4)      & 8.2(1)           &          &6.2(4)                              \\ 
            ZrNi$_3$          &  76(1)      &  8.5(1)       &          & 4.3(6)                   \\    
          Zr$_8$Ni$_{21}$               &   98.2(5)   &  11.0(2)         &          & 6.5(3)                     \\    
            Zr$_7$Ni$_{10}$           &  64(2)         &  7.3(1)     & 2.3(3)         &                   \\ \\

HfNi$_3^{(1)}$                   & 32.9(4)         & 3.7(1)   &1.6(2)           &                         \\   
     HfNi$_3^{(2)}$                   & 65.7(4)     & 7.3(1)          &                & 7.0(4)                        \\ 
            Hf$_2$Ni$_7$                & 72.2(4)   &8.1(1)             &               & 5.7(2)                        \\   
   Hf$_8$Ni$_{21}$                    & 99.3(6)  &11.1(2)              &                &6.6(4)                       \\  
      Hf                   & 53.3(6)            & 6.0(1)  &0.8(2)          &                       \\  \\
\hline
\end{tabular}}
\label{tab:ZrNi3_HfNi3_fitting_table}
\end{center}
\end{table}

\begin{figure*}[t!]
\begin{center}
\includegraphics[width=0.75\textwidth]{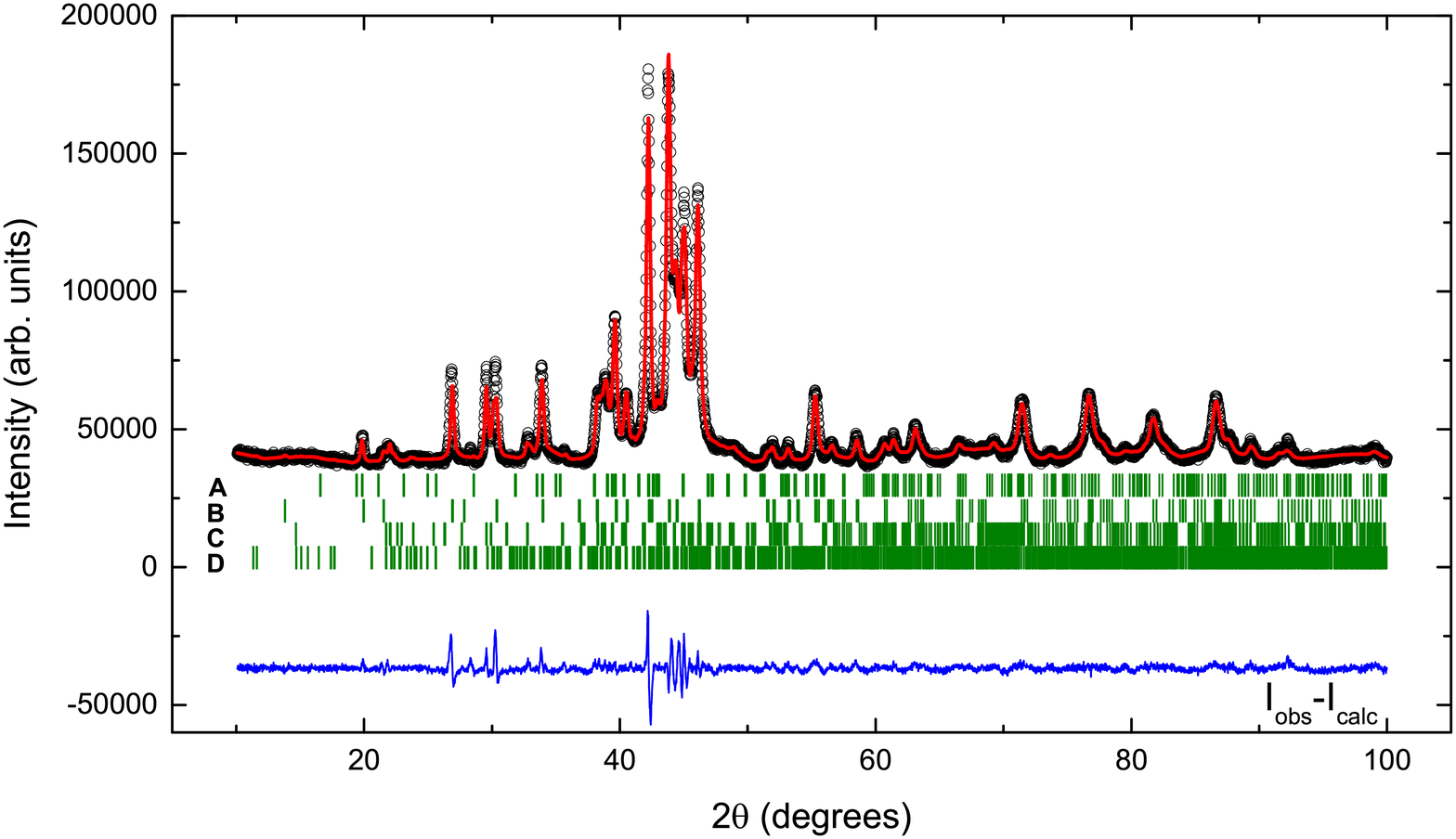}
\end{center}
\captionof{figure}{\small{ The background subtracted XRD powder pattern in HfNi$_3$. The line represents the fit to the measured data. 
The vertical bars A, B, C and D denote the Bragg angles corresponding to $\beta$-HfNi$_3$, $\alpha$-HfNi$_3$, Hf$_2$Ni$_7$ and Hf$_8$Ni$_{21}$, respectively.
The bottom line shows the difference between the observed and the fitted pattern.}}
\label{HfNi3_XRD}
\end{figure*}

\begin{figure*}[t!]
\begin{center}
\includegraphics[width=0.9\textwidth]{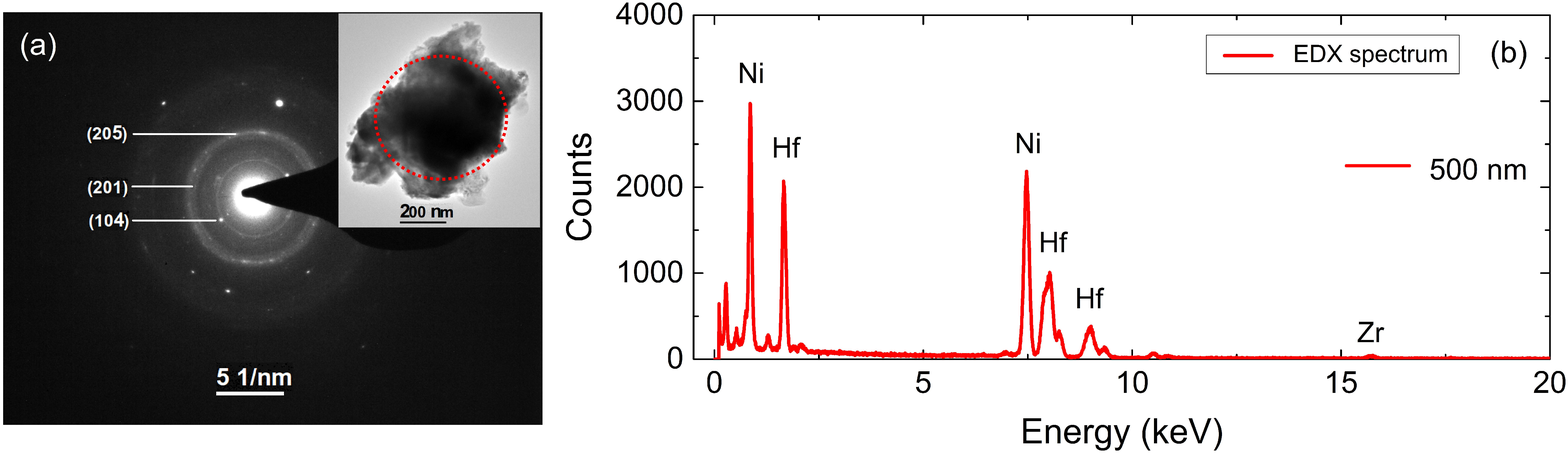}
\end{center}
\captionof{figure}{\small{(a) Selected area electron diffraction pattern from $\beta$-HfNi$_3$ particle shown in the inset. (b) Energy dispersive X-ray spectrum from
the same particle.}}
\label{HfNi3_TEM}
\end{figure*}

\begin{figure*}[t!]
\begin{center}
\includegraphics[width=0.7\textwidth]{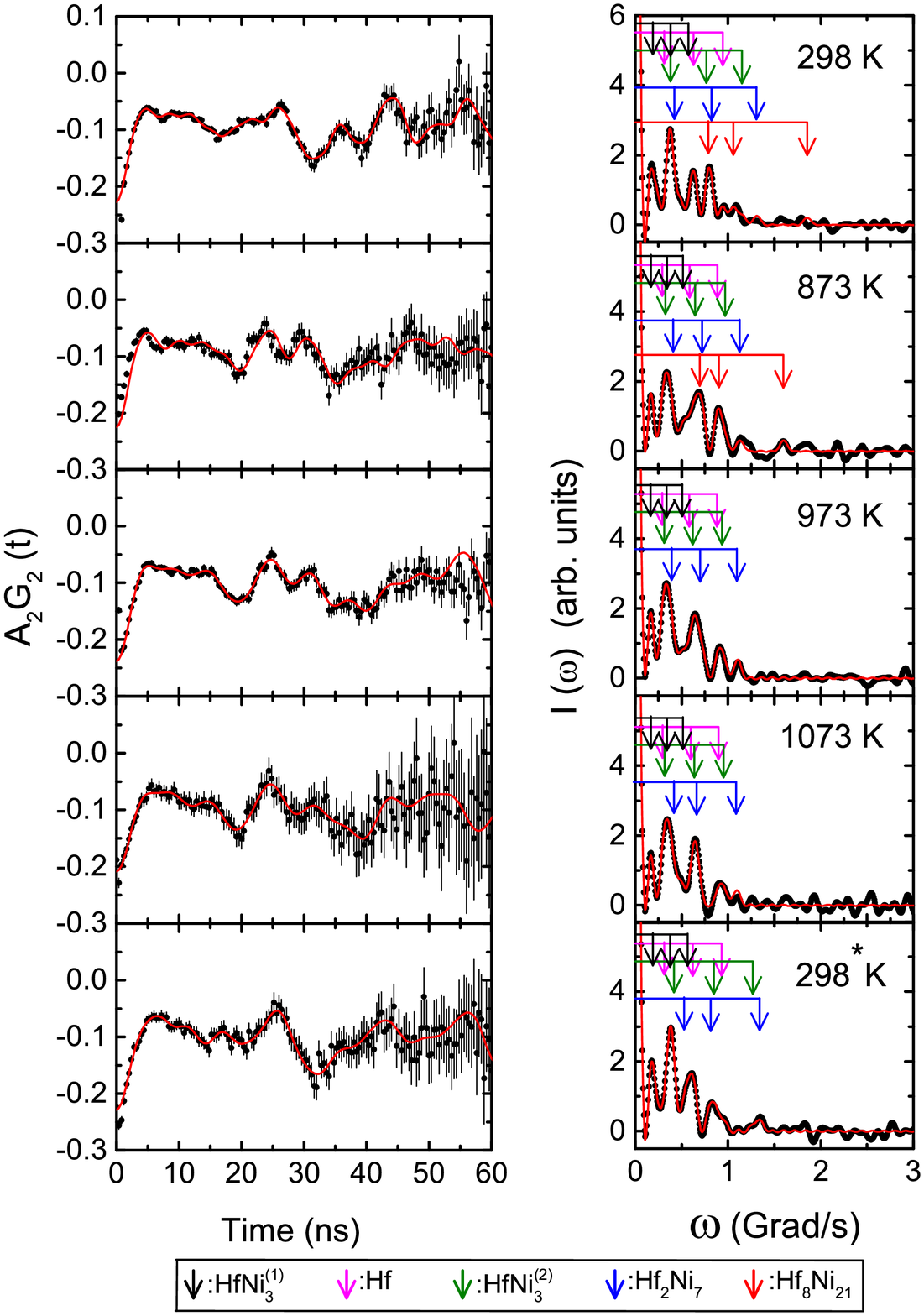}
\end{center}
\captionof{figure}{\small{TDPAC spectra in the stoichiometric sample of HfNi$_3$ at different
temperature. Left panel
shows the time spectra and the right panel shows
the corresponding Fourier cosine transforms. The PAC spectrum at room temperature designated by 298$^*$ K is taken after the measurement at 1073 K. A set of three
arrows indicates the three transition frequencies of a particular component.
}}
\label{HfNi3_spectra}
\end{figure*}

\begin{figure}[t!]
\begin{center}
\includegraphics[width=0.5\textwidth]{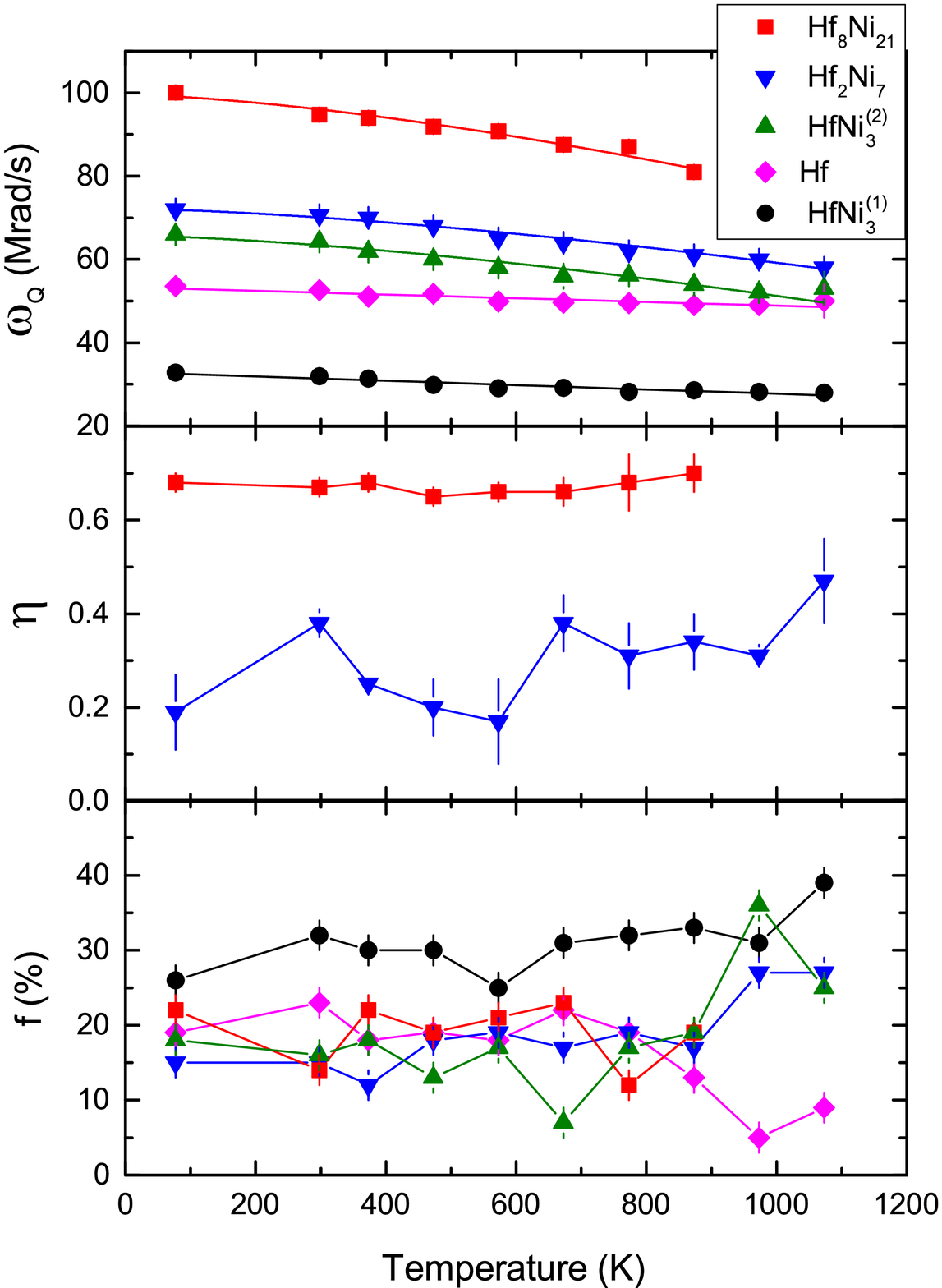}
\end{center}
\captionof{figure}{\small{Variations of quadrupole frequency ($\omega_Q$), asymmetry parameter ($\eta$) and site fraction $f$(\%) with
temperature for HfNi$_3^{(1)}$, HfNi$_3^{(2)}$, Hf$_2$Ni$_7$, Hf$_8$Ni$_{21}$ and Hf.}}
\label{HfNi3_parameter}
\end{figure}

 \section{PAC Results}
 
 \subsection{Stoichiometric ZrNi$_3$ sample}
 The X-ray powder diffraction pattern found in the stoichiometric sample of ZrNi$_{3}$ is shown in Figure \ref{ZrNi3_XRD}. The peaks in
the spectrum were identified using ICDD database, 2009. Presence of Zr$_2$Ni$_7$ [PDF Card No.: 01-071-0543], ZrNi$_3$ [PDF Card No.: 01-029-0946],
Zr$_8$Ni$_{21}$ [PDF Card No.: 01-071-2622] and Zr$_7$Ni$_{10}$ [PDF Card No.: 03-066-0045] phases have been found in the XRD spectrum. Analysis of the X-ray powder pattern was
carried out by FULLPROF software package \citep{Rodriguez} using the known crystallographic parameters 
of Zr$_2$Ni$_7$ \citep{Eshelman}, ZrNi$_3$ \citep{Becle}, Zr$_8$Ni$_{21}$ \citep{JM} and Zr$_7$Ni$_{10}$ \citep{Yvon}. From present XRD 
analysis, refined
values of lattice parameters obtained are shown in Table \ref{tab:HfNi3_ZrNi3_structure}. The presence of ZrNi$_3$ phase in
this stoichiometric sample of ZrNi$_3$ has been observed from TEM/EDX measurement (Figure \ref{ZrNi3_TEM}) also. The atomic percentages for Zr and Ni at the indicated spot have been found 
to be 24.4(3) and 75.6(2), respectively. Selected area electron diffraction (SAED) pattern obtained from a region marked by a dotted circle for the
same particle in the stoichiometric sample of ZrNi$_3$ is shown in Fig. \ref{ZrNi3_TEM}. Some of the measured interplaner spacings ($d$-spacing) from the SAED pattern
are 2.07(4) \r{A}, 2.86(4) \r{A} and 3.60(4) \r{A}. These measured $d$-spacings are very close to the (201), (110) and (101) inter-planer
spacing of hexagonal ZrNi$_3$ (JCPDS $\#$ 29-0946), respectively. This further confirms the presence of ZrNi$_3$ phase in the sample.
  
The PAC spectrum in the stoichiometric sample of ZrNi$_3$ at room temperature is shown in Figure \ref{ZrNi3_spectra}. The spectrum was best
fitted by considering
four electric quadrupole interactions. The sample produced was found to have non-random orientation of microcrystals and the spectrum was fitted by 
considering free $S_{2n}$ coefficients. The results of different components found are shown in Figure \ref{ZrNi3_parameter}. 
The main frequency component ($\sim$48\%) produces values of $\omega_Q$=70.9(2) Mrad/s, $\eta$=0.25(2). 
This component can be assigned to 
Zr$_2$Ni$_7$ by comparing the values of $\omega_Q$ and $\eta$ with the earlier reported results in Zr$_2$Ni$_7$ \citep{Marszalek,CCDey_physica}. From previous PAC
measurements in ZrNi$_5$ also, a similar component to this was obtained and attributed to Zr$_2$Ni$_7$ \citep{Silva, CCDeyMag}. The ZrNi$_5$ has a cubic crystal 
structure and no EFG at the probe site is expected due to ZrNi$_5$.
The component 2 with a 
symmetric EFG ($\eta\sim$0) can be attributed to ZrNi$_3$ by comparing with our calculated results from density functional theory (discussed later).
The crystal structure of ZrNi$_3$ is hexagonal close-packed and, therefore, a value of 
$\eta$=0 is expected for this compound. However, this is found to be
a minor phase compared to other phases produced in this sample.
The results of component 3 can be compared with our recent results in Zr$_8$Ni$_{21}$ \citep{skdeyZr8Ni21}. 
This component is found to be similar to one component of Zr$_8$Ni$_{21}$ found from our previous measurements \citep{skdeyZr8Ni21} and can,
therefore, be attributed to Zr$_8$Ni$_{21}$. The
component 4 can be attributed to Zr$_7$Ni$_{10}$ by comparing with the results found in Zr$_7$Ni$_{10}$ \citep{skdeyZr7Ni10}. A similar 
component to this was found in Zr$_8$Ni$_{21}$ also where it was attributed to Zr$_7$Ni$_{10}$ \citep{skdeyZr8Ni21}.

The results of temperature dependent PAC measurements in the stoichiometric ZrNi$_3$ are shown in the Figure \ref{ZrNi3_parameter}.
The corresponding TDPAC spectra are shown in Figure \ref{ZrNi3_spectra}. It is found that the component fraction due to ZrNi$_3$ is present in the temperature 
range (77-973 K). At 77 K, the site fraction of ZrNi$_3$ was found to be maximum ($\sim$30\%). In the 
temperature range 77-873 K, four component fractions are found to be present with no appreciable change in parameters. The fractional variations of different components are shown in Figure \ref{ZrNi3_parameter}.
At 973 K, a distinct 
change 
in PAC spectrum has been observed. At this temperature, the component due to Zr$_7$Ni$_{10}$ disappears. On the other hand, a new frequency component
with values of 
$\omega_Q$=53.0(4) Mrad/s, $\eta$=0 appears. This component is similar to that found in Zr$_8$Ni$_{21}$ at 1073 K \citep{skdeyZr8Ni21}.
The site percentage of this new component is found to be $\sim$39\% at this temperature and it enhances abruptly at 1073 K ($\sim$90\%). At
1073 K, only two 
components are found to be present. The minor 
component found at this temperature is due to Zr$_2$Ni$_7$. At this temperature, ZrNi$_3$ and Zr$_8$Ni$_{21}$ phases completely disappear. We have repeated the
measurement at room temperature after the 
measurement at 1073 K. The remeasured spectrum at room temperature produces  
a strong electric quadrupole interaction ($\sim$81\%) with values of $\omega_Q$=56.5(1) Mrad/s, $\eta$=0. This component can be recognized as
the same component that appeared at 973 and 1073 K. Besides this, two other minor components are found here. The component due to Zr$_8$Ni$_{21}$ 
reappears with a small fraction ($\sim$8\%) and the component due to Zr$_2$Ni$_7$ is also found to be present ($\sim$10\%). No component 
of ZrNi$_3$ at room temperature after heating the sample to 1073 K has been observed. This indicates a decomposition of ZrNi$_3$ at 1073 K. 

From previous XRD measurements \citep{Bulyk}, ZrNi$_3$ was found to be produced after heating ZrCrNi-H$_2$ system to 1024 K and a
decomposition of ZrNi$_3$ 
to Zr$_2$Ni$_7$ was observed \citep{Bulyk} after heating the sample to 1083 K. The results of present PAC measurements, therefore, support
the results of previous XRD measurements \citep{Bulyk}. On the other hand, decomposition of ZrNi$_3$ at 1213 K as reported by Becle et al. \citep{Becle} 
is not supported.

In the present Zr-Ni sample, the predominant component found at 1073 K and subsequently at room temperature can probably be assigned to Hf. 
At temperatures below
973 K, the probe
atoms which were settled at various lattice sites come out from the lattice positions after gaining sufficient energy at high temperature. It 
seems that at 973 K, probe atoms have detached partially ($\sim$39\%) from the Zr-Ni compounds and at 1073 K, 
only a small fraction of the probe nucleus ($\sim$10\%) is attached with the compounds. A 
similar phenomenon was 
observed from our recent PAC investigation in Zr$_8$Ni$_{21}$ \citep{skdeyZr8Ni21}. 

The evolution of quadrupole frequency, asymmetry parameter and site fraction with temperature for different components observed 
are shown in Figure \ref{ZrNi3_parameter}. It is found that quadrupole frequencies for the components Zr$_2$Ni$_7$, ZrNi$_3$ and
Zr$_8$Ni$_{21}$
vary with temperature following
$T^{3/2}$ relationship. For the  Zr$_8$Ni$_{21}$ component, a similar temperature dependent behavior was observed from our previous PAC investigation in
Zr$_8$Ni$_{21}$ \citep{skdeyZr8Ni21}. For these three components, values of $\omega_Q$ have been fitted using the relation
  \begin{equation}
   \omega_Q(T)=\omega_Q(0)[1-\beta T^{3/2}] 
   \label{eqn:T32}
  \end{equation}
  where, $\omega_Q(0)$ is the extrapolated value at 0 K.
  The results of $\eta$ and site fraction for different components are also plotted (Figure \ref{ZrNi3_parameter}). These results do not show large 
  variations except the site fraction of Zr$_2$Ni$_7$ which decreases drastically at 1073 K ($\sim$10\%) compared to the fraction found at 973 K ($\sim$41\%). 
The fitted results are listed in Table \ref{tab:ZrNi3_HfNi3_fitting_table}. Contrary to these, $\omega_Q$ for the component Zr$_7$Ni$_{10}$
is found to obey a linear temperature 
dependent behavior. In this case, we have fitted the results of $\omega_Q(T)$ using the relation 
  \begin{equation}
   \omega_Q(T)=\omega_Q(0)[1-\alpha T]. 
   \label{eqn:T}
  \end{equation}
Both $T^{3/2}$ and $T$ variations of EFG (proportional to $\omega_Q$) for metallic and intermetallic systems are found in
literature \citep{Christiansen}. 

\begin{table}[t!]
\begin{center} 
\captionof{table}{\small{The parameters of the $\beta$-HfNi$_3$ and ZrNi$_3$ structure, given in \r{A}.}}
\scalebox{0.67}{
\begin{tabular}{ l l l l l } 
 \hline  \\ [-0.9ex]
&Our calculated results   & Previous experimental  & Earlier
calculated  & Present experimental \\ &(WIEN 2k) \citep{Blaha,Murnaghan} &results (X-ray diffraction) \citep{HfNi3Structure,Becle} &results \citep{Anubhav} & results (X-ray diffraction) \\ [1.5ex]
 \hline  \\

$\beta$-HfNi$_3$ & & \\	
$a$ &5.285 &5.2822(2) &5.267 &5.282(1)\\
$c$ &21.419 &21.3916(18) &21.411 &21.392(2)\\
$B$ [GPa] &186 & & \\  
Hf 2b &0 0 1/4 &0 0 1/4 &0 0 1/4 \\
Hf2 4f &1/3 2/3 0.3488 & 1/3 2/3 0.3488 &1/3 2/3 0.3489 \\
Hf3 4f &1/3 2/3 0.5458 &1/3 2/3 0.5458 &1/3 2/3 0.5461 \\
Ni 6h &0.5110 0.022 1/4 &0.5117 0.0234 1/4 & 0.5107 0.0213 1/4 \\
Ni2 12k &0.156 0.312 0.0514 & 0.156 0.312 0.0514 & 0.1563 0.3126 0.0512 \\
Ni3 12k &0.8320 0.6640 0.1495 &0.8316 0.6632 0.1495 & 0.8322 0.6645 0.1496 \\ \\
 ZrNi$_3$ & & \\
$a$ &5.319 &5.3090(8) &5.267 &5.308(2)\\
$c$ &4.305 &4.3034(12) &21.411 &4.303(1)\\
$B$ [GPa] &177 & &\\  
Zr 2c &1/3 2/3 1/4 &1/3 2/3 1/4 &1/3 2/3 1/4\\ 
Ni 6h &0.8435 0.687 1/4 &0.829 0.658 1/4 &0.844291 0.688581 1/4 \\ \\ 
\hline
\end{tabular}}
\label{tab:HfNi3_ZrNi3_structure}
\end{center}
\end{table}

\begin{figure*}[t]
\begin{center}
\includegraphics[width=0.8\textwidth]{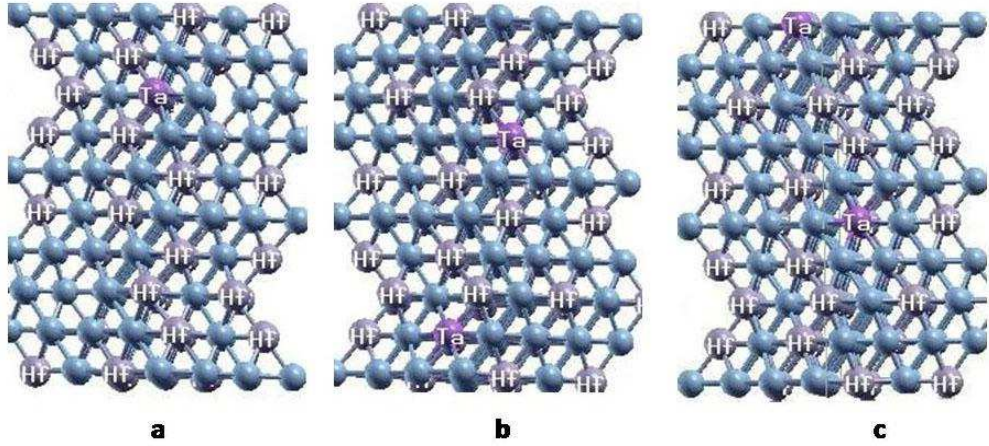}
\end{center}
\captionof{figure}{\small{Models of cells used in the study of HfNi$_3$}}
\label{HfNi3_crystal}
\end{figure*}

\begin{figure}[t]
\begin{center}
\includegraphics[width=0.4\textwidth]{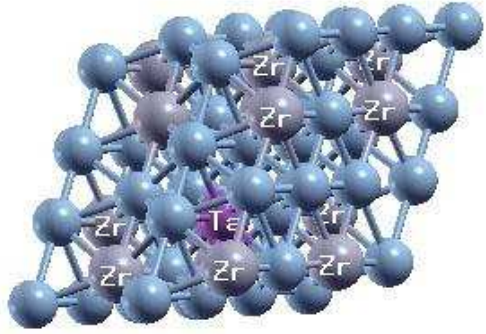}
\end{center}
\captionof{figure}{\small{Model of cell used in the study of ZrNi$_3$}}
\label{ZrNi3_crystal}
\end{figure}

\begin{table}[t!]
\begin{center} 
\captionof{table}{\small{  Calculated EFG values for HfNi$_3$ in units of 10$^{21}$ V/m$^2$ and asymmetry parameters}}
\scalebox{0.75}{
\begin{tabular}{ l l l l } 
 \hline  \\ [-0.9ex]
Probe  &Lattice Site    & EFG  & asymmetry \\&&& parameter ($\eta$)  \\ [1.5ex]
 \hline  \\ 
 
no probe           &Hf1 2b 0 0 1/4 &-1.3 &0 \\
    (pure compound)                            &Hf2 4f 1/3 2/3 0.3488(2) &-5.0 &0 \\
                              &Hf3 4f 1/3 2/3 0.5458(1)  &-2.2 &0 \\ \\

$^{181}$Ta    &Hf1 2b 0 0 1/4 &-1.7 &0 \\
                        &Hf2 4f 1/3 2/3 0.3488(2) &-7.1 &0 \\
                        &Hf3 4f 1/3 2/3 0.5458(1)   &-3.5 &0  \\ \\                       
                                              
\hline
\end{tabular}}
\label{tab:DFT_HfNi3}
\end{center}
\end{table}
\subsection{Stoichiometric HfNi$_3$ sample}
The XRD measurement in the stoichiometric sample of HfNi$_{3}$ has been carried out at room temperature (Figure \ref{HfNi3_XRD}). The peaks in
the spectrum were identified using ICDD database, 2009. Presence of $\beta$-HfNi$_3$ [PDF Card No.: 01-071-0476], $\alpha$-HfNi$_3$ [PDF Card No.: 01-071-0474],
Hf$_2$Ni$_7$ [PDF Card No.: 01-074-6880] and Hf$_8$Ni$_{21}$ [PDF Card No.: 01-074-0476] phases have been found in the XRD spectrum. Analysis of the X-ray powder pattern was
carried out by FULLPROF software package \citep{Rodriguez} using the known crystallographic data 
of $\beta$-HfNi$_3$ \citep{HfNi3Structure}, $\alpha$-HfNi$_3$ \citep{HfNi3Structure}, 
Hf$_2$Ni$_7$ \citep{Dattagupta} and Hf$_8$Ni$_{21}$ \citep{LBsenko}. From present XRD analysis, refined values of lattice parameters found
are shown in
Table \ref{tab:HfNi3_ZrNi3_structure}. It was reported \citep{Bsenko} that high temperature phase of HfNi$_3$ was
formed from the melt and Hf$_2$Ni$_7$ by peritectic reaction. Apart from HfNi$_3$, there are phases due to Hf$_2$Ni$_7$ and Hf$_8$Ni$_{21}$. But, no prominent 
peak due to Hf$_7$Ni$_{10}$ was found in the XRD spectrum. The phase HfNi$_3$ has been observed from our TEM/EDX measurement also (Figure \ref{HfNi3_TEM}). The atomic
percentages for Hf and Ni at the indicated spot have been found to be 
24.4(2) and 74.3(2), respectively. The SAED pattern obtained from a region marked by a dotted circle for the
same particle in the stoichiometric sample of HfNi$_3$ is shown
in Fig. \ref{HfNi3_TEM}. Some of the measured interplaner spacings ($d$-spacing) from the SAED pattern
are 2.06(4) \r{A}, 2.31(4) \r{A} and 3.41(4) \r{A}. These measured $d$-spacings are very close to the (205), (201) and (104) inter-planer
spacing of hexagonal $\beta$-HfNi$_3$ (JCPDS $\#$ 71-0475), respectively. This further confirms the presence of $\beta$-HfNi$_3$ phase in the sample.

The PAC spectrum observed in HfNi$_3$ at room temperature after preparing the sample in argon arc furnace is shown in Figure \ref{HfNi3_spectra}.
It is found that five electric quadrupole
frequencies are required to fit the time spectrum. Analysis was done by considering free $S_{2n}$ coefficients. The major component ($\sim$32\%) found with 
values of $\omega_Q$=32.0(3) Mrad/s, $\eta$=0 can be attributed to HfNi$_3$ by comparing with our calculated results from
DFT (discussed later). From
our previous studies in Hf$_8$Ni$_{21}$ \citep{skdeyZr8Ni21} and Hf$_7$Ni$_{10}$ \citep{skdeyZr7Ni10}, a component similar to this was observed and tentatively
assigned to HfNi$_3$. Component 2 can also be attributed to HfNi$_3$ because
the values of EFG and 
asymmetry parameter for this component are found to be in good agreement with the calculated results from DFT (discussed later). The component 3 
has been attributed to Hf$_2$Ni$_7$ by comparing with the earlier 
reported results in Hf$_2$Ni$_7$ \citep{Marszalek, Silva}. From previous PAC measurement in HfNi$_5$, a similar component to this was also found and attributed
to Hf$_2$Ni$_7$ \citep{Silva}. The component 4 ($\sim$14\%) with values of $\omega_Q$=94.8(6) Mrad/s and $\eta$=0.67(2) can be assigned to Hf$_8$Ni$_{21}$. From our recent
investigation in Hf$_8$Ni$_{21}$ \citep{skdeyZr8Ni21}, similar
values of quadrupole 
frequency and asymmetry parameter were found. In Hf$_8$Ni$_{21}$, however, two non-equivalent Hf sites were found. But, in this stoichiometric HfNi$_3$ sample,
we have
found only one site of Hf$_8$Ni$_{21}$. The other non-equivalent site of Hf$_8$Ni$_{21}$ is not observed here. Besides these, a component with symmetric EFG 
(component 5) was found which can be attributed to pure hcp Hf by comparing the values of $\omega_Q$ and $\eta$
with earlier reported results \citep{skdeyHfmetal}. Probably, 
this component arises due to unreacted Hf with Ni. Decrease of this component at 973 and 1073 K indicates
that more Hf reacts with Ni to form compounds at high temperatures.  

Temperature dependent PAC results are shown in Figure \ref{HfNi3_parameter}. The corresponding PAC spectra are shown in Figure \ref{HfNi3_spectra}. 
In the temperature range 77-873 K, there are no appreciable changes in the PAC spectra. At 973 K, the component Hf$_8$Ni$_{21}$ disappears. 
The component due to Hf$_8$Ni$_{21}$ does not appear at 1073 K also. But other four components are found to exist at 1073 K. However, unlike stoichiometric
ZrNi$_3$, no additional component is observed at 1073 K. The PAC measurement was then repeated at room temperature. 
At this temperature, two components of HfNi$_3$ reappear which indicates that HfNi$_3$ is a stable phase. The component Hf$_8$Ni$_{21}$ does not appear when remeasured at room temperature.
The components of Hf$_2$Ni$_7$ and Hf are
found to be present when remeasured at room temperature.  

The evolution of the quadrupole frequency, $\eta$ and site fraction with temperature for the different components observed in stoichiometric HfNi$_3$ are shown 
in Figure \ref{HfNi3_parameter}. The components Hf$_8$Ni$_{21}$ and Hf$_2$Ni$_7$ follow the same $T^{3/2}$ temperature dependence 
as found 
for Zr$_8$Ni$_{21}$ and Zr$_2$Ni$_7$ in ZrNi$_3$. The EFG for the two components of HfNi$_3$ are found to vary different manner. The quadrupole frequency of HfNi$_3^{(2)}$ varies with temperature following $T^{3/2}$
relationship (Eqn. \ref{eqn:T32}).
On the other hand, a linear temperature dependent behavior (Eqn. \ref{eqn:T}) was found for the HfNi$_3^{(1)}$
component. 
Variation of
the quadrupole frequency for the hexagonal Hf was also found to be linear. The fitted results are shown in Table \ref{tab:ZrNi3_HfNi3_fitting_table}. 
The variations of $\eta$ and site fractions for different components do not show large changes (Figure \ref{HfNi3_parameter}).

\section{DFT calculations}

The first-principles density functional theory (DFT) calculations were
performed with the WIEN2k simulation package \citep{Blaha} based on the full potential
(linearized) augmented plane waves method (FP (L)APW). Electronic exchange-correlation
energy was treated with generalized gradient approximation (GGA) parametrized by
Perdew-Burke-Ernzerhof (PBE) \citep{Perdew}. In our calculations the muffin-tin radii for Hf,
Ni, Zr and Ta were 2.3, 2.1, 2.3 and 2.3 a. u., respectively. The cut-off parameter $R_{mt}K_{max}$
for limiting the number of plane waves was set to 7.0, where $R_{mt}$ is the smallest
value of all atomic sphere radii and $K_{max}$ is the largest reciprocal lattice vector
used in the plane wave expansion.

The Brillouin zone integrations within the self-consistency cycles were performed
via a tetrahedron method \citep{Blochl}, using 6-50 $k$ points in the irreducible wedge of the
Brillouin zone (4$\times$4$\times$2 and 8$\times$8$\times$8 meshes for Ta doped HfNi$_3$ and ZrNi$_3$, respectively) for the supercell calculations. The atomic
positions were relaxed according to Hellmann-Feynman forces calculated at
the end of each self-consistent cycle, with the force minimization criterion
2 mRy/a.u.. In our calculations the self-consistency was achieved by demanding
the convergence of the integrated charge difference between last two iterations
to be smaller than 10$^{-5}$ $e$. All the calculations refer to zero temperature.
\subsection{HfNi$_3$}
HfNi$_3$ at the temperatures below 1200$^\circ$C, has the $\gamma$-Ta(Pd,Rh)$_3$-type structure,
with a stacking of ten AB$_3$ layers in the sequence ABCBCACBCB. The space group
is $P6_3/mmc$ and the unit cell dimensions are $a$=5.2822(2) \r{A}, $c$=21.3916(18) \r{A} at room temperature \citep{HfNi3Structure}. This structure contains 40 atoms in the unit cell,
distributed at 6 non-equivalent crystallographic positions, 3 for Hf atoms
and 3 for Ni atoms (Table \ref{tab:HfNi3_ZrNi3_structure}).

After obtaining the optimized structural parameters, we constructed 2$\times$2$\times$1 supercell
from periodically repeating unit cells of the host crystals. To simulate PAC measurements
at Hf1 position, we replaced one Hf atom in the supercell at the position (0 0 1/4) with Ta (Figure \ref{HfNi3_crystal}a \citep{Kokalj}).
In the case of Ta at the Hf2 and Hf3 positions due to the complexity of the calculations, we had to
replace two Hf atoms at the corresponding position with Ta, thus obtaining the cell with 50 non-equivalent atoms (Figure \ref{HfNi3_crystal}b and c). We checked
that the two Ta atoms are sufficiently far from each other (11.1 \r{A}) to avoid significant
impurity-impurity interactions. 
After determing the self-consistent charge density we obtain the electric field gradient
(EFG) tensor $V_{ij}$ using the method developed in reference \citep{Schwarz}. The usual convention is to designate
the largest component of the EFG tensor as $V_{zz}$. The asymmetry parameter $\eta$ is then given
by $\eta$= ($V_{xx}$-$V_{yy}$)/$V_{zz}$, where $|V_{zz}|\ge|V_{yy}|\ge|V_{xx}|$. All the calculations refer to zero temperature.

The theoretically determined cell and structure parameters for the
investigated structure, along with the experimental values
obtained from X-ray diffraction measurements
are given in Table \ref{tab:HfNi3_ZrNi3_structure}. The theoretical volume slightly
overestimates the experimental one. The bulk modulus $B_0$,
obtained by fitting the data to the Murnaghan's equation
of state \citep{Murnaghan} is also given in Table \ref{tab:HfNi3_ZrNi3_structure}. The calculated formation
enthalpy, -0.50 eV/atom, is in good agreement with the earlier measured (-0.52 eV/atom \citep{deBoer}) and
calculated (-0.44 eV/atom \citep{Johannesson}, -0.54 eV/atom \citep{Levy}) values. 

The calculated EFGs in the pure compound as well as at
Ta probe position in the $\beta$-HfNi$_3$ are given in Table \ref{tab:DFT_HfNi3}.
It can be observed that EFG is smallest at Hf1 position
and the largest at Hf2 position. This trend preserves also
for the electric field gradients calculated at corresponding
Ta positions, but the EFGs are now larger from 30\% to 60\%. We see that
the calculated result for EFG at the Ta probe site replacing Hf3
atom (3.5$\times$10$^{21}$ V/m$^2$) is in excellent agreement with
the measured value of  EFG=3.7(1)$\times$10$^{21}$ V/m$^2$ ($\omega_Q$(0)=32.9(4) Mrad/s) for the component HfNi$_3^{(1)}$, thus confirming that
the mentioned component of the
measuered PAC spectra originates from HfNi$_3$. Similarly, the calculated results at the Ta probe site replacing
Hf2 atom (V$_{zz}$=7.1$\times$10$^{21}$ V/m$^2$ and $\eta$=0) are in excellent aggrement with our measured values of HfNi$_3^{(2)}$ component
(V$_{zz}$(0)=7.3(1)$\times$10$^{21}$ V/m$^2$ and $\eta$=0) which confirms that this component also originates from HfNi$_3$.

\subsection{ZrNi$_3$}
ZrNi$_3$ crystallizes in the hexagonal Ni$_3$Sn type structure,
which possesses two non-equivalent crystallographic positions,
Zr 2c and Ni 6h \citep{Becle}. The optimized lattice constants, which slightly
overestimate the experimental values, are given in Table \ref{tab:HfNi3_ZrNi3_structure}. The calculated formation
enthalpy, -0.41 eV/atom, agrees well with the earlier
calculated (-0.36 eV/atom \citep{Johannesson}, -0.46 eV/atom \citep{Hennig}) values.
The calculated EFG at Zr position is -3.0$\times$10$^{21}$ V/m$^2$, with
zero asymmetry parameter. In order to simulate PAC measurement,
we constructed 2$\times$2$\times$2 supercell from periodically repeating unit
cell and then replaced one of the Zr atoms by Ta (Figure \ref{ZrNi3_crystal} \citep{Kokalj}). The point group
symmetry around the impurity Ta atom remained the same as around
the original Zr atom, but the number of non-equivalent positions
increased. The calculated EFG at the Ta probe atom -8.4$\times$10$^{21}$ V/m$^2$
is in excellent agreement with two mutually similar EFG values
from measured PAC spectra (8.2 and 8.48$\times$10$^{21}$ V/m$^2$, corresponding
to $\omega_Q$(0)= 72.9 and 76 Mrad/s, respectively). The fact that
the corresponding calculated asymmetry parameter is zero, enables
us to assign the 76 Mrad/s component to ZrNi$_3$ and thus definitely
confirm the presence of this phase in our stoichiometric sample.

\section{Conclusion} 
From TDPAC and XRD measurements, multiple phases have been found in the stoichiometric samples of ZrNi$_3$ and HfNi$_3$. The presence of ZrNi$_3$ and 
HfNi$_3$ in these stoichiometric samples have been confirmed from TDPAC, XRD and TEM/EDX measurements. From PAC studies, it is
found that ZrNi$_3$ is produced as a minor phase while the phase HfNi$_3$ is found to be largely produced. Also, our temperature dependent PAC studies
show that HfNi$_3$ is a very stable phase. In the stoichiometric samples of ZrNi$_{3}$ and HfNi$_{3}$, 
secondary phases due to (Zr/Hf)$_8$Ni$_{21}$ and (Zr/Hf)$_2$Ni$_7$ are found to be produced. In ZrNi$_3$, the phase
due to Zr$_7$Ni$_{10}$ is observed while no phase due to Hf$_7$Ni$_{10}$ is found in
HfNi$_3$ sample. Only one 
and the same crystallographic site of (Zr/Hf)$_8$Ni$_{21}$ is found in present stoichiometric samples of ZrNi$_3$ and HfNi$_3$ although two non-equivalent
sites were found in
(Zr/Hf)$_8$Ni$_{21}$ \citep{skdeyZr8Ni21}. The experimental values of EFG and $\eta$ for ZrNi$_3$ and HfNi$_3$ are found to be in excellent agreement with
the theoretically calculated 
values of EFG and $\eta$ at $^{181}$Ta impurity sites by the first-principles density functional theory based on the FP (L)APW. From our calculation,
three non-equivalent Hf sites in HfNi$_3$ have been found whereas two of these have been observed from PAC measurements. In ZrNi$_3$, on the other hand, the
present DFT calculation produces one EFG corresponding to a single Zr site. From our PAC measurements in ZrNi$_3$ also, a single frequency component has been found.

The solubility of Hf in Ni is found to be less compared to Zr in Ni. The Hf solubility in Ni is found to increase
with temperature 
and it decreases again when the temperature is lowered. In Zr-Ni compounds, the binding energy of Hf probe to the lattice
sites is not strong enough at high temperature and, probably, the probe atoms are detached from the 
compound at 1073 K. 

The TDPAC is found to be an useful nuclear technique to detect weak component phases that are produced in a material. Particularly, a
LaBr$_3$(Ce)-BaF$_2$ set up is found to be very useful for separating the minor component phases when multiple components are present in the sample.

\vspace{1 cm}

 {\hspace{ -0.4 cm}}{\bf Acknowledgement}
  \vspace{0.5cm}
  
 The help of Prof. Dr. T. Butz, University of Leipzig, Germany in data
analysis is gratefully acknowledged with thanks. We are grateful to Dr. B. Satpati of SINP, Kolkata for TEM/EDX measurements and analysis. We would
like to thank A. Karmahapatra and S. Pakhira of
SINP, Kolkata for their helps in XRD
measurements and data analysis. The present work is supported by the Department of Atomic Energy, Government of India through the Grant
no. 12-R\&D-SIN-5.02-0102. J. Belo$\check{\text{s}}$evi\'{c}-$\check{\text{C}}$avor and D. Toprek acknowledge support by The Ministry of Education, 
Science and Technological Department of the Republic of Serbia through the Grant no. 171001.

\end{document}